\documentclass{emulateapj}

\def\sun{\odot}
\def\earth{\oplus}

\begin{document}

\title{Using Kuiper Belt Binaries to Constrain Neptune's Migration History}
\author{Ruth A.~Murray-Clay\altaffilmark{1} and Hilke
E.~Schlichting\altaffilmark{2,3}} \altaffiltext{1}{Harvard-Smithsonian Center
for Astrophysics, 60 Garden Street, MS-51, Cambridge, MA 02138, USA}
\altaffiltext{2} {UCLA, Department of Earth and Space Science, 595 Charles E.
Young Drive East, Los Angeles, CA 90095, USA}\altaffiltext{3} {Hubble
Fellow}\email{rmurray-clay@cfa.harvard.edu}

\begin{abstract}
Approximately 10--20\% of all Kuiper belt objects (KBOs) occupy mean-motion resonances with Neptune. This dynamical configuration likely resulted from resonance capture as Neptune migrated outward during the late stages of planet formation. The details of Neptune's planetesimal-driven migration, including its radial extent and the concurrent eccentricity evolution of the planet, are the subject of considerable debate.  Two qualitatively different proposals for resonance capture have been proposed---migration-induced capture driven by smooth outward evolution of Neptune's orbit and chaotic capture driven by damping of the planet's eccentricity near its current semi-major axis.  We demonstrate that the distribution of comparable-mass, wide-separation binaries occupying resonant orbits can differentiate between these two scenarios. If migration-induced capture occurred, this fraction records information about the formation locations of different populations of KBOs.  Chaotic capture, in contrast, randomizes the orbits of bodies as they are placed in resonance.   In particular, if KBO binaries formed by dynamical capture in a protoplanetary disk with a surface mass density typical of observed extrasolar disks, then migration-induced capture produces the following signatures.  The 2:1 resonance should contain a dynamically cold component, with inclinations less than 5--10$^\circ$, having a binary fraction comparable to that among cold classical KBOs.  If the 3:2 resonance also hosts a cold component, its binary fraction should be 20--30\% lower than in the cold classical belt.  Among cold 2:1 (and if present 3:2) KBOs, objects with eccentricities $e < 0.2$ should have a binary fraction $\sim$20\% larger than those with $e > 0.2$.  Other binary formation scenarios and disk surface density profiles can generate analogous signatures but produce quantitatively different results.  Searches for cold components in the binary fractions of resonant KBOs are currently practical. The additional migration-generated trends described here may be distinguished with objects discovered by LSST.
\end{abstract}

\section{INTRODUCTION}\label{sec-intro}

Our Kuiper belt, a $\sim$0.01--0.1$M_\oplus$ \citep[e.g.,][]{BTA04,
clm+07, FH108} collection of planetesimal debris located beyond the
orbit of Neptune, provides a unique window into the dynamical processes that
shaped the young solar system.  The orbits of Kuiper belt objects (KBOs)
comprise a dynamical fossil, recording the movements of the giant planets
during the era of their formation.

Though almost two decades have passed since the discovery of the first KBO
after Pluto and Charon \citep{jl93}, and though the orbits of more than 500
trans-Neptunian objects have been well-characterized
\citep[e.g.,][]{ekc+05,kjg+09}, a detailed theoretical
understanding of the Kuiper belt's rich dynamical structure remains elusive.
For our purposes, we will focus on the following aspects of this structure,
which remain difficult to reconcile.
\begin{itemize}
\item Approximately 10--20\% of all KBOs, including Pluto and Charon, occupy mean
motion resonances with Neptune \citep{jlt98, TJL01, kjg+09}, and this fraction may be larger if many objects reside in distant resonances which are not yet well-characterized.  In
resonance, the orbital period of a KBO forms an approximate integer ratio with
that of Neptune, leading to a coherent exchange of energy and angular momentum
with the planet which is periodic over $\sim$$10^4$--$10^5$ year timescales.  Though resonance occupation enhances long-term stability \citep[e.g.,][]{lm05}, the large number of resonant KBOs suggests a special dynamical origin for this population.
\item Several lines of evidence suggest that classical KBOs, those objects that are not in resonance and are not currently undergoing close encounters with Neptune \citep[e.g.,][]{gmv08}, consist of two populations with different histories and having different typical inclinations.  The inclination
distribution of the belt is bimodal, best fit by a population of bodies with $i \lesssim
5^\circ$ and a distinct population containing equal to greater mass
with inclinations ranging up to
$\sim$$35^\circ$ \citep{b01,ekc+05,gea+10}.  In addition, low and high inclination classical KBOs have systematically different physical
properties.  The high inclination population is bluer in color
\citep{tr00,tb02,PL08}, has a lower fraction of wide binaries \citep{SN06,ngs+08}, and contains all of the brightest (and hence likely largest) KBOs \citep{ls01a}.  The critical inclination separating those objects with bluer and redder colors is $\sim$$10^\circ$ rather than $\sim$$5^\circ$ \citep{lm05,PL08}.  The significance of this discrepancy, if any, is not yet known and may be complicated by overlap of the low and high inclination populations.  Throughout
this paper, we refer to this entire population as the classical belt, the low
inclination subset as the cold classicals and the high inclination subset as
the hot classicals.  When specific numbers are used, we define ``low inclination" as those objects with $i < 10^\circ$ and ``high inclination" as those with $i \ge 10^\circ$.  We choose $10^\circ$ as our cutoff based on the distributions of colors with inclination because we will be investigating a physical property of KBOs---their binary fraction.  Our conclusions will not change substantially if $5^\circ$ is determined to be more appropriate for the classical belt.
\item Classical KBOs exist with eccentricities ranging up to $\sim$0.2.  However, cold classical KBOs have systematically low eccentricities, likely
reflecting a dynamically unexcited population.  We note that for the purposes of this paper, ``cold" and ``hot" refer only to inclination.  
\item The inclinations of KBOs in 3:2 mean motion resonance span the same
range as the inclinations of the classical belt, but the dynamical data may be fit with a single high
inclination population \citep{gea+10}.
\item At $\sim$48 AU, the density of cold classical KBOs declines precipitously \citep{tb01,abm02}.  This transition is located near
Neptune's 2:1 mean-motion resonance.  Though the presence of the edge is
clear, whether it is an edge in semi-major axis corresponding to the 2:1
resonance is less certain due to observational selection effects at these
large distances from the Sun.  Further, this edge may not be present in the hot classical population (Petit et al., in prep).  
\end{itemize}
No dynamical model has yet simultaneously reproduced the inclination
distribution of the belt, its eccentricity distribution, and the resonant
population of KBOs.

Two qualitatively different explanations have been proposed for the abundance
of KBOs occupying mean-motion resonances with Neptune.  The first, which we refer to as ``migration-induced capture," was proposed
by \citet{m93,m95} to explain the orbit of Pluto. In this model, KBOs are captured into
resonance as Neptune migrates outward during the late stages of planet
formation.  This migration is powered by planetesimal scattering in a disk containing $\sim$$30 M_\oplus$ in planetesimals
\citep{fi84,gml04}, which also acts to damp the eccentricity and inclination
of Neptune.  In the simplest version of this scenario, Neptune migrates
$\sim$10 AU with low eccentricity and inclination, though it may well have
occupied a more excited orbit at earlier times.  As this migration progresses,
Neptune entrains KBOs in its mean motion resonances as the semi-major axes of
the resonances sweep through the primordial planetesimal disk \citep{m93,m95}.
This primordial disk must contain some KBOs with pre-excited eccentricities \citep{cjm+03}
to explain the occupation of high-order
resonances such as the 5:2.  
In this context, \citet{lm03} argue that
the entire Kuiper belt might have been pushed out from closer to the Sun by
migration, citing the belt's low mass and external edge near Neptune's 2:1
resonance.

As explored by \citet{hm05}, early outward migration of Neptune reproduces the
resonant structure of the belt well, but suffers from two difficulties.
First, resonance capture excites the eccentricities of captured particles, but
does not change their inclinations substantially, so inclinations must be
excited prior to migration or an alternative inclination excitation mechanism
must be found.  \citet{g03} argue that the large inclination objects in the
belt may have arisen from scattered objects which interacted with secular
resonances during migration.  This process has a low efficiency
\citep{clm+07}, however, making it difficult to explain the number of large
inclination objects.  Second, smooth migration does not sufficiently excite
the eccentricities of non-resonant objects to explain the hot classical
population of KBOs.  \citet{lm03} argue that stochastic effects during
migration could cause KBOs with excited eccentricities to be lost from
resonance, populating the hot classical belt.  However, \citet{mc06}
demonstrate that for a substantial number of KBOs to be lost from resonance
during migration, more than a few percent of the mass in the planetesimal disk
driving migration must be in bodies with radii larger than
$\mathcal{O}(1000{\rm km})$, comparable to the size of Pluto.  This mass
fraction is inconsistent with estimates of the size distribution in the early
disk resulting from coagulation \citep[e.g.,][]{kl99,SR11}.  Whether this
conclusion is altered by recent models of planetesimal formation that may
allow rapid formation of 100--1000 km bodies \citep{yg05,jom+07} remains to be
seen.

A third difficulty for migration-induced capture arises from the
different inclination distributions in the classical Kuiper belt and the 3:2
resonant population.  Resonant capture does not substantially alter inclinations, implying that the inclination distribution of a resonant population should match that of the population from which it captured its members.  Though the 3:2 resonance did not pass through the region of the currently observed classical belt, it captured its population from an adjacent region of the protoplanetary disk, where one would naively expect the inclination distribution to be the same.  Nevertheless, it is possible that objects caught into the 3:2 resonance experienced additional dynamical sculpting, rendering this expectation invalid, though no model for such sculpting as been produced thus far.  In particular, we note that if distinct ``hot" and ``cold" populations exist in the 3:2 resonance, the characteristic inclination dividing these populations may not match that in the classical belt.  The possibility of additional sculpting is less plausible for the 2:1 resonance, which did pass through the classical belt.  Unfortunately, the inclination distribution of the 2:1 is not yet constrained well enough to confirm or rule out a cold component.

A second possibility for the population of mean-motion resonances was
suggested more recently, inspired by the Nice model for the
early evolution of the solar system giants \citep{tgm+05}.  In this model, Jupiter and Saturn cross their mutual 2:1 mean motion resonance as a result of migration driven by planetesimal scattering.  This resonance passage substantially alters the orbits of Uranus and Neptune.  \citet{lmv+08} argue that it is
difficult to place Neptune interior to 21 AU at the end of the Nice model, and
suggest chaotic population of the Kuiper belt as an alternative to
migration-induced resonance capture.  They argue that Neptune was scattered,
with large eccentricity, onto an orbit with a semi-major axis close to its
current value of 30 AU.  Conservation of angular momentum during the
scattering event leaves the planet on a low-inclination orbit.  Initially
after the scattering, orbits with semi-major axes ranging from that of Neptune
out to its 2:1 resonance are chaotic.  This chaos results from overlapping
mean-motion resonances which are rendered wide by Neptune's large
eccentricity.  As Neptune's eccentricity is damped by its interactions with
planetesimals, the resonances decrease in width, ending in their current
configuration.  Objects in mean motion resonances are more likely to remain
stable during this process than their non-resonant counterparts, particularly
at high eccentricities.  We refer to this process as ``chaotic capture."

We note that this model does have a final phase of slow migration of Neptune
on a low-eccentricity, low-inclination orbit, however its extent is less than
a few AU.   \citet{lmv+08} argue that the high inclinations achieved in their
simulations result from this late migration phase and the \citet{g03}
mechanism.  The inclination distribution of the Kuiper belt is not
well-matched in this or any model, leaving the source of the high inclination
population of the Kuiper belt an important outstanding problem.  
Chaotic capture is appealing due to its natural explanation for
the edge of the belt coincident with Neptune's 2:1 mean-motion resonance.  It
does not, however, reproduce the low-inclination, low-eccentricity classical
Kuiper belt.  Furthermore, an {\it in situ} belt having the properties of the classical
KBOs is excited too much by Neptune while the planet's orbit is eccentric.

In this paper, we present a new method for distinguishing between competing
models of Kuiper belt sculpting based on observations of the binary fraction
of KBOs. A significant fraction of large KBOs exist in binaries. More than 70
such systems are currently known and this number will continue to rise as a result of new KBO
searches including Pan-STARRS and LSST.  \citet{ngs+08} have shown that
the fraction of KBOs that are binaries
varies with dynamical class in the Kuiper belt. In particular, cold classical
KBOs (defined in \citet{ngs+08} to have $i<5.5~\rm{deg}$) have a binary fraction of $\sim 29\%$, while hot
classical KBOs have a fraction of only $\sim 10\%$. Furthermore, the physical characteristics of binaries in the two populations differ.  All of the new
binary systems reported by \citet{ngs+08} with low heliocentric inclinations are similar brightness systems and therefore presumably
consist of roughly equal mass companions. On the contrary, among the binary
systems in the hot population, there were both similar brightness systems and
systems with large brightness differences between the binary
components.   \citet{SN06} found that the rate of binaries for the hot
classical, resonant, and scattered populations combined is about 5.5\%.  Binary statistics are not yet sufficient to establish the binary fraction in the resonances alone.  

We
hypothesize that this observed variation in binary fraction arises ultimately from
variations in the location in the protosolar nebula at which binary formation
took place. 
Broadly speaking, one can identify two classes of Kuiper belt
binaries. The first class is comprised of the large KBOs that are orbited by
small satellites. The second class consists of similar brightness systems with
typically wide separations. This first class of systems probably originated
from a collision and tidal evolution, as it has been proposed for the
formation of the Moon and the Pluto-Charon system \citep{HD75,CW76,MK89}. Such
a formation scenario fails however for the second class of Kuiper belt
binaries, since it cannot account for the large angular momenta (i.e. wide
separations) of these systems. 

Motivated by this challenge a series of new binary formation scenarios, described in \S\ref{sec-bincap}, were proposed.  In many of these scenarios, binaries form by dynamical capture processes that operated before the Sun's planetesimal disk was excited by the giant planets.  Though dynamical capture and subsequent orbital evolution can produce binaries with small separations, wide-separation, roughly equal mass systems stand out as products of dynamical capture rather than collisions, and we focus on these systems.  In most dynamical capture models, the rate at which binaries form varies with heliocentric distance.  

We will demonstrate that migration-induced capture of KBOs into mean-motion
resonances with Neptune generates an observable variation with orbital
properties in the fraction of KBOs that are wide-separation, roughly equal-mass binaries.  This variation arises because a current KBO's orbit is correlated with the location in the protoplanetary disk at which that KBO formed.  In
particular, we will demonstrate that low-inclination resonant Kuiper belt
objects could retain a measurable trend in binary fraction with heliocentric
eccentricity.  We calculate this
trend and argue that if observed, it will provide evidence both that the cold
classical population of the belt formed near its current location and that
Neptune experienced a period of smooth migration.  In contrast, if resonant KBOs were emplaced by
chaotic capture as Neptune's eccentricity damped, then resonant KBOs were dynamically mixed after excitation of the disk halted binary capture, and the binary fraction should not correlate with eccentricity.  
For narrative clarity, we discuss the
migration-induced capture mechanism for emplacing low inclination resonant KBOs for the majority of this paper, and we return
to a comparison with chaotic capture at the end.  This choice should not
be interpreted as favoring one model above the other.  Rather, we hope that a
future census of Kuiper belt binaries will provide a useful discriminant
between these classes of models.  

Binary fraction is not the only, or even the first, physical property of KBOs that may be searched for dynamical signatures.  In particular, no trend in color with eccentricity has been observed for resonant KBOs \citep[e.g.,][]{dbt+08}.  However, for dynamical studies, binary fraction has the substantial advantage over color that there exists a quantitative theory for how it likely varied within the young Sun's planetesimal disk.  Consequently, we focus on binary fraction throughout the bulk of this paper.  We return to a discussion of KBO colors and sizes in Section \ref{sec-colors}.

In Section \ref{sec-stage}, we provide a schematic model of migration-induced
resonance capture.  Section \ref{sec-dynam} describes the dynamical history of
a KBO captured into resonance during a long Neptunian migration.  Section \ref{sec-bincap} reviews how binary planetesimals,
which will ultimately become binary KBOs, form.  We calculate formation rates
as a function of distance from the Sun, and we emphasize that this formation
must occur before the belt is substantially excited or depleted. We combine
these results to predict the fraction of wide-separation, comparable-mass KBO binaries as a function of dynamical
class in Section \ref{sec-binfrac}, and we argue that these trends should not be
present if resonances were populated via chaotic capture.  Given current projections, these trends will be measurable using KBOs discovered by LSST.  We consider our test in the context of  established trends in KBO colors and sizes (Section \ref{sec-colors}), summarize, and conclude (Section \ref{sec-sum}).

\section{FRAMEWORK}\label{sec-stage}

In this section, we assert a schematic of the dynamical origin of resonant and classical Kuiper belt objects, inspired by migration-induced capture.  We will use this model as a framework for the majority of this paper.  We emphasize both that this is not a full dynamical model and that we are not advocating this scenario as the true history of the outer solar system.  Rather, we will demonstrate that dynamical models which share the broad characteristics of our schematic generate a unique observable signature in binary KBO dynamics which will be absent in other types of models including chaotic capture.  Thus, it presents a useful framework for interpreting observations.

In our scenario, planetesimals interior to $\sim$$30$ AU from the Sun have their eccentricities and inclinations excited by the giant planets after the planets themselves undergo a dynamical excitation event.  The prevalence of extrasolar planets with large eccentricities \citep[e.g.,][]{bwm+06} suggests that such events are common, and most current models of planet formation include dynamical upheavals, whether they result from resonance crossings as in the Nice model \citep{tgm+05} or from reductions in the efficiency of damping by small bodies and gas \citep{GLS042}.  Beyond $\sim$30 AU, a planetesimal disk formed and survived with low eccentricity and inclination.  After this period of dynamical excitation, the eccentricity and inclination of Neptune were damped by the residual planetesimal disk at a semi-major axis of $\sim$20--25 AU.  It then migrated outward to 30 AU, maintaining low eccentricity and inclination, as it scattered planetesimal debris \citep{fi84}.  

Whether the above scenario can reproduce, in detail, the eccentricity and inclination properties of the belt remains to be determined and is the subject of ongoing research.  For our purposes, the relevant characteristics of a model such as this are as follows.  The cold classical belt, with low inclinations, formed {\it in situ}.  High inclination KBOs formed closer to the Sun than low inclination KBOs.   The high inclination and low inclination populations of the Kuiper belt were generated by distinct dynamical mechanisms, have distinct histories, and were formed at different locations in the disk.  Furthermore, high inclination KBOs experienced more dynamical mixing than low inclination KBOs.  As Neptune migrated out through the planetesimal disk, it captured both high and low inclination objects into resonance.  Low inclination objects were likely to be captured near their formation locations, while high inclination objects may have been moved substantially before capture.

Even in smooth migration scenarios, the hot population of classical KBOs is likely to be dynamically mixed from its formation location and may be subject to enhanced binary disruption due to scattering by Neptune.  Given these considerations, we suggest that the searches for coherent signatures of migration in the physical properties of KBOs should be restricted to the low inclination population.

\section{ORBITAL PROPERTIES OF RESONANT KBOS CAPTURED VIA MIGRATION}\label{sec-dynam}

The final eccentricity $e_{\rm fin}$ of an object caught into $p$:$q$ resonance by a slowly migrating Neptune can be related to the semi-major axis at which it was captured, $a_{\rm cap}$, using an adiabatic invariant of the system.  After reviewing relevant properties of resonances, we describe this invariant, originally derived for a system containing a single planet on a circular orbit, interacting with a test particle occupying the same plane (Section \ref{sec-orbprelim}), demonstrate that it is useful under a range of likely histories for the outer solar system (Section \ref{sec-varymig}), and comment on the relationship between eccentricity and inclination among observed resonant KBOs (Section \ref{sec-evsi}).

\subsection{Orbital Preliminaries and Brouwer's Constant}\label{sec-orbprelim}

First, we briefly review pertinent facts about mean motion resonances.  Further discussion may be found in, e.g., the textbook \citet{md00}.  Consider a system consisting only of a central star with mass $M_*$, a planet with mass $M_p$ on a circular orbit, and a collection of test particles.  A particle in external $p$:$q$ mean motion resonance with the planet experiences libration of the resonance angle
\begin{equation}\label{eqn-phi}
\phi = p\lambda - q\lambda_p - (p-q)\varpi \;\; ,
\end{equation}
where $\lambda$ and $\lambda_p$ are the mean longitudes of the test particle and the planet, respectively, and $\varpi$ is the longitude of pericenter of the particle.  This expression is appropriate for resonances dominated by the eccentricity of the particle, as is the case for all known resonant KBOs.  For a first-order resonance ($p - q = 1$), the angle $\phi$ is approximately the angular position of the test particle with respect to pericenter at the time of conjunction, when the planet lies on the line connecting the particle to the star.  Because the gravity of the planet generates the largest perturbation to the particle's orbit at conjunction, if $\phi$ remains constant, then the planetary perturbations coherently add.  This resonant effect can generate large changes to the particle's orbit.  
%Higher order resonances are weaker because they correspond to patterns of $p-q$ angles of conjunction---at $(\phi + 2\pi k)/(p-q)$ for integer $k$---offering greater opportunities for the perturbations at conjunction to partially cancel.  

When resonance is not exact, $\phi$ is not a constant, but rather librates through a limited range of values.  The timescale for this libration is $t_{\rm lib} \approx C_1 t_{\rm orb} [(M_*/M_p)e^{-|p-q|}]^{1/2}$, where the particle has orbital period $t_{\rm orb} = 2\pi/\Omega$, orbital angular frequency $\Omega$, and eccentricity $e$.  Libration within the resonance produces variations in the particle's semi-major axis, $a$, with a maximum magnitude (full width) of $\max(\delta a_{\rm lib}) \approx 2C_2 a [(M_p/M_*)e^{|p-q|}]^{1/2}$.
The coefficients $C_1$  and $C_2$ are order unity constants.  For the 3:2 resonance, $C_1 \approx 0.3$ and $C_2 \approx 3.6$.  For the 2:1, $C_1 \approx 0.9$ and $C_2 \approx 3.0$. 

Now imagine that the planet's orbit expands, while remaining circular.  As long as this outward migration is sufficiently slow, the orbit of a particle in external mean motion resonance also expands, maintaining the resonant lock.  As it does so, its eccentricity evolves such that
\begin{equation}\label{eqn-aeinvar}
C_B = a(\sqrt{1-e^2}-q/p)^2   \;\;,
\end{equation}
where $C_B$, known as Brouwer's constant, is an adiabatic invariant of the system \citep{b63,yt99,hm05}.  This relationship applies as long as the particle remains in resonance and the timescale of migration $t_{\rm mig} \equiv a_p/\dot a_p \gg t_{\rm syn}$, the synodic period (time between conjunctions).  In general, the former is a stricter constraint.  To capture and maintain particles in resonance, the planet must migrate slowly enough that $[\max(\delta a_{\rm lib})/a] t_{\rm mig} \gg t_{\rm lib}$.

Fundamentally, Brouwer's constant encodes the ratio between angular momentum and energy transfer during transport in resonance.  For a resonant particle, the specific energy $E = -GM_*/(2a)$ and angular momentum, $L = (GM_*a(1-e^2))^{1/2}$, measured with respect to the host star, increase at rates $\dot E$ and $\dot L$, respectively, such that 
\begin{equation}\label{eqn-emom}
\dot E \approx \dot L\Omega_p
\end{equation}
when averaged over a synodic period \citep[e.g.,][]{clm+07}.  In contrast, maintenance of a circular orbit would
require $\dot E = \dot L\Omega$.  Equation (\ref{eqn-emom}) may be intuitively
understood as the result of forcing at the planet's orbital frequency $\Omega_p$.  The
radius $\vec r$ and the velocity $\vec v$ of a Keplerian orbit are roughly
perpendicular to one another, and their magnitudes differ by approximately
$\Omega$, so one might expect a perturbing force $\vec F$ to generate a torque $\vec r
\times \vec F$ and rate of work $\vec F \cdot \vec v$ that differ by
approximately $\Omega$.  This intuition fails because
the majority of the
energy and angular momentum change averages to zero over a synodic period.  Only second-order effects remain, which instead
yield a term proportional to the forcing frequency $\Omega_p$.

In this paper, we appeal to Brouwer's constant to relate the semi-major axis at
which a particle was caught into resonance during Neptune's migration to its
current eccentricity in resonance.  For example, a straightforward application
of Equation (\ref{eqn-aeinvar}) implies that under the hypothesis of smooth
migration, a KBO currently having $e_{\rm fin}=0.2$ and semi-major axis $a_{\rm fin} = 39.5$ AU in Neptune's 3:2
resonance was captured at $a_{\rm cap} \approx 34.9$ AU if it initially had zero
eccentricity.  This consideration is the basis for estimates of the extent of
Neptune's primordial migration---if KBOs were all caught with low eccentricities, the
current eccentricities of resonant objects imply that Neptune began its migration at
least 10 AU closer to the Sun than its present orbit \citep{m93,hm05}.  In
Section \ref{sec-varymig}, we discuss the accuracy of this mapping in the
context of the solar system.

\subsection{Conservation of Brouwer's Constant Under Realistic Migration Conditions}\label{sec-varymig}

Given the assumption that smooth migration of a low-eccentricity Neptune
occurred, one may still worry that difficulties will arise in applying Brouwer's
constant, developed for the circular restricted three-body problem, to the
true solar system.  The signal might be washed out due to the presence of four
giant planets, all with small but notable eccentricities and inclinations, by
the fact that we do not, {\it a priori}, know the initial eccentricities of
KBOs, or by chaotic evolution over the full age of the solar system.
Fortunately, this is not the case.

To evaluate our ability to use Brouwer's constant to identify the semi-major axis at which a KBO was captured into resonance, we performed a numerical integration
of KBOs evolving under the influence of migrating planets.  We performed this integration in two steps, both using the hybrid package of the N-body integrator Mercury, version 6.2 \citep{c99}.  In the first step, we begin with 10,000 test particles with initial eccentricities ranging uniformly between 0 and 0.2 and initial semi-major axes ranging uniformly between 25 and 50 AU.  We include all four giant planets with their current eccentricities and inclinations.  The initial semi-major axes of the planets differ from their current values by $\Delta a = 7$, 3, 0.8, and $-0.2$ AU for Neptune, Uranus, Saturn, and Jupiter, respectively, and they evolve to their current 
semi-major axes over an exponential timescale $\tau = 10^7$ as in \citet{m95}.  This migration is implemented by applying accelerations to the giant planets of the form $0.5(\dot a/a) {\bf v} $ in the Mercury6 routine {\verb mfo_user }, where $\dot a/a = (\Delta a/a)\tau^{-1} \exp(-t/\tau)$ at time $t$ and ${\bf v}$ is the vector velocity of the particle.  This integration lasts for $3\times10^7$ years with a time step of 8 days.  At the end of $3\times10^7$ years, we continue the integration with no migration force and with a time step of 150 days for a further $10^9$ years only for the 1350 objects that were captured into 3:2 (647) or 2:1 (703) resonance with Neptune.  Because we do not employ the migration force over the entire integration, the planets end with semi-major axes that differ from their true values by approximately $ \exp(-3)\Delta a \approx 0.05 \Delta a$.  After $10^9$ years, 65 objects remain in the 3:2 resonance and 47 in the 2:1.  We note that this corresponds to a retention efficiency smaller than the 39\% and 24\% retention efficiencies calculated over 1 Gyr by \citet{tm09} for the 3:2 and 2:1 resonances, respectively, due to chaotic diffusion under the influence of the four giant planets.  This difference likely results either from differences in the population of resonant objects at the beginning of our long term integrations or from the slightly different configuration of the giant planets in our long-term integration.  Our integrations are intended to verify our ability to retrieve initial semi-major axes under a wide range of initial conditions rather than to faithfully reproduce the initial conditions in the disk, so we do not consider our substantially lower retention fraction significant.

\begin{figure}[ht]
\vspace{-0.5in}
%\plotone{aicompare_long.pdf}
%\plotone{f1.eps}
%\plotone{f1.pdf}
%\epsscale{0.5}
\includegraphics[scale=0.6]{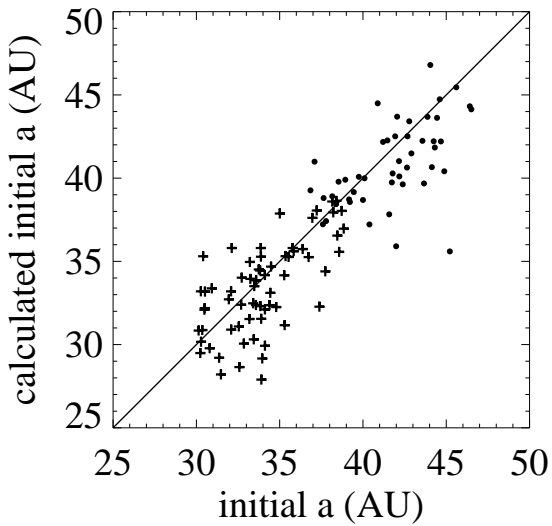}
\caption{True initial semi-major axis of 3:2 (plusses) and 2:1 (circles) KBOs captured into
resonance via migration in a numerical simulation compared against the semi-major axis at the time of capture inferred from Equation (\ref{eqn-ai}).  The integration includes all four
giant planets with their current eccentricities and inclinations.  Test
particles have initial eccentricities ranging from 0 to
0.2.  This trend may be substantially tightened by consideration of a second adiabatic invariant of the system (Murray-Clay, Lithwick, \& Chiang, in prep).}\label{fig-aicomp}
\end{figure}

Figure \ref{fig-aicomp} compares the true
initial semi-major axis with the initial semi-major axis $a_{\rm cap}$ calculated
using Equation (\ref{eqn-aeinvar}) for the KBOs which were captured and maintained in 3:2 or 2:1 resonance for the full $10^9$ years.  In order to apply Equation (\ref{eqn-aeinvar}) without {\it a priori} knowledge of the initial orbits of the particles, we assume zero initial
eccentricity so that
\begin{equation}\label{eqn-ai}
a_{\rm cap} = a_{\rm fin} \frac{(\sqrt{1-e_{\rm fin}^2}-q/p)^2}{(1-q/p)^2} \;\;.
\end{equation}
Figure \ref{fig-aicomp} exhibits a clear trend between the true initial $a$
and the calculated value.  The scatter in this relation corresponds to a cut at $e = 0.2$ misclassifying the initial semi-major axis of approximately 20\% of observed objects.  This scatter can be substantially reduced by
evaluation of a second adiabatic invariant of the system which permits us to
estimate the initial eccentricity of a captured KBO, allowing for a more accurate
calculation of the initial semi-major axis of a captured particle
(Murray-Clay, Lithwick, \& Chiang, in prep.).  However, this improved
calculation is substantially more complicated than application of Equation
(\ref{eqn-aeinvar}) and is beyond the scope of this paper.  We have verified
that our ability to recover the initial semi-major axis is not lost after
evolution over the age of the solar system.

Since all particles in a given resonance have comparable semi-major axes,
Equation (\ref{eqn-ai}) implies that particles caught into resonance at
smaller semi-major axes have larger final eccentricities and that resonant
evolution can lead to substantial eccentricity excitation.  This excitation
was one of the original motivations for the idea of Neptune's migration
\citep{m93}. 

\subsection{Eccentricity vs.~Inclination}\label{sec-evsi}

\begin{figure}[ht]
%\plotone{eiplot.pdf}
%\plotone{f2.eps}
%\plotone{f2.pdf}
%\epsscale{0.5}
\vspace{-0.2in}
\includegraphics[scale=0.45]{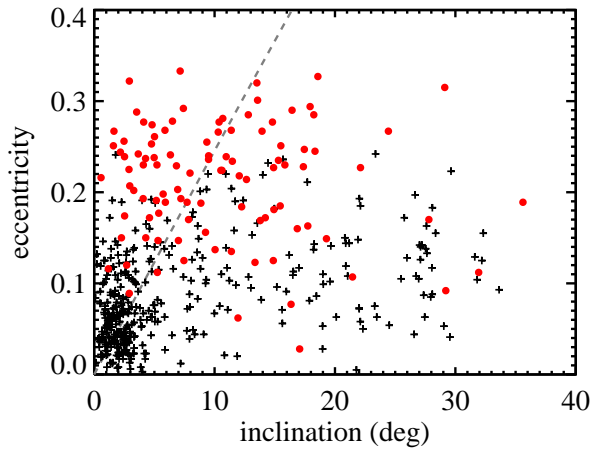}
\caption{Eccentricity vs.~inclination with respect to the invariable plane
for observed classical KBOs (black crosses) and 3:2 resonant KBOs (red circles) from the MPC catalog.  The eccentricities of
3:2 resonant KBOs are systematically higher than the eccentricities of classical KBOs and do not correlate with inclination.  This pattern is consistent with
migration-induced eccentricity excitation, which does not affect inclination.
Because 3:2 objects are more stable than classical bodies at high eccentricity, this pattern is also consistent with removal of high $e$ classical objects due to planet scattering, as occurs, for example, in chaotic capture.  
We include for reference the line $e = \sqrt{2}\sin(i)$, representing equipartition excitation of eccentricity and inclination.}
\label{fig-eiclass}
\end{figure}

We note that differences in the relationship between eccentricity and inclination, $i$,
for non-resonant and resonant objects are broadly consistent with
migration-induced capture.  
Figure \ref{fig-eiclass} displays the relationship between $e$ and $i$ for known classical and 3:2 resonant KBOs in the Minor Planet Center (MPC) catalog\footnote{{\tt http://www.cfa.harvard.edu/iau/lists/TNOs.html}, March 31, 2010.} with at least 3 observed oppositions.  Inclinations are calculated with respect to the invariable plane of the solar system.  The invariable plane differs from the true dynamical plane of the Kuiper belt---known as the forced or Laplace plane---by at most a few degrees \citep{bp04, cc08}.  We use the dynamical classifications of \citet{gmv08}, which includes objects with 3 oppositions as of May 2006.  For the approximately 20\% of each population of objects not contained in that work, we assign a tentative dynamical classification based on a $10^5$ year integration of the nominal orbit in the presence of the four giant planets.\footnote{Objects for which the 3:2 resonance angle $\phi$ librates are assigned to the 3:2 resonance. An object is designated classical if its semi-major axis $a > 30$ AU, its eccentricity $e < 0.24$, it survives the $10^5$ year integration, $\phi$ does not librate for the 5:4, 4:3, 3:2, 5:3, 7:4, 9:5, 2:1, 7:3, or 5:2 resonances, and finally either its pericenter $q = a(1-e) > 35$ AU or the Tisserand parameter $T = a_N/a + 2\sqrt{(a/a_N)(1-e^2)}\cos(\Delta i) > 3$, where $a_N$ is the semi-major axis of Neptune and $\Delta i$ is the mutual inclination between the planes of Neptune and the KBO.}

As mentioned in Section \ref{sec-intro}, 3:2 objects do not exhibit the same
excess of low inclination bodies seen in the classical population.  This lack
of a large cold population is potentially problematic in the context of
migration-induced capture, but is not conclusive because the 3:2 resonance did
not sweep through the region of the classical belt.  However, among hot
objects, the distribution of eccentricity and inclination mimics that expected
for migration-induced capture.  The inclinations of KBOs in the 3:2 resonance
are not correlated with eccentricity and span the same range as the
inclinations of hot classical objects.  At the same time, the eccentricities
of the resonant KBOs are systematically higher than those of the classical
objects.  This pattern of similar inclinations but offset eccentricities is
consistent with migration-induced capture, which excites eccentricities while
leaving inclinations largely unchanged.  This pattern is suggestive but not
conclusive evidence for migration-induced capture because high eccentricity
KBOs are more stable in the 3:2 resonance than in the classical belt, and the
observed eccentricity offset could alternatively result from the removal of
high $e$ classical bodies.

\section{BINARY FORMATION AS A FUNCTION OF SEMI-MAJOR AXIS}\label{sec-bincap}

Having established that under the migration-induced capture scenario, we can statistically recover the the initial semi-major axes of KBOs in mean motion resonance with Neptune, we now show that the binary fractions of different populations of KBOs are expected to be a function of the locations in the protoplanetary disk at which those populations formed.  In Section \ref{sec-binform}, we review proposed binary formation scenarios.  We argue that binary formation mediated by dynamical friction with small bodies, as proposed by \citet{GLS02}, likely formed the Kuiper belt binaries.  In Section \ref{sec-dfmech}, we review the \citet{GLS02} formation mechanism, and in Section \ref{sec-abund}, we calculate the
binary fraction as a function of formation distance.  We note that other binary formation mechanisms which vary with distance could also generate an observable, though different, signature in the resonant population.

\subsection{Binary Formation Scenarios}\label{sec-binform}

High mass-ratio binaries in the Kuiper belt, including Pluto/Charon, likely formed in collisions \citep{MK89}.  However, the majority of observed binary KBOs are roughly equal-mass, wide-separation binaries \citep{ngs+08} that have too much angular momentum to be formed by the same mechanism.  Motivated by this challenge, a series of new
binary formation scenarios was proposed
\citep[e.g.][]{W02,GLS02,F04,A05,laf07,nyr10}.  Most of these scenarios appeal to interactions within the Hill sphere, the region interior to the Hill radius of a KBO.
The Hill radius denotes the distance from a KBO at which the tidal forces from
the Sun and the gravitational force from the KBO, both acting on a test
particle, are in equilibrium. It is given by
\begin{equation}\label{e1}
R_{H} \equiv a \left( \frac{m}{3 M_*}\right) ^{1/3} 
\end{equation}
where $a$ is the semi-major axis of the KBO and $m$ its mass.  \citet{W02}
proposed a collision between two KBOs inside the Hill sphere of a third.
However, in the Kuiper belt, gravitational scattering between the two
intruders is about 100 times\footnote{For this estimate we used that
$\alpha\sim R/R_H\sim 10^{-4}$ and assumed that the velocity dispersion of the
KBOs at the time of binary formation is less than their Hill velocity (see \S
\ref{sec-dfmech}).} more common than a collision. Binary formation by three
body gravitational deflection ($L^3$ mechanism), as proposed by \citet{GLS02},
should therefore dominate over such a collisional formation scenario. A second
binary formation scenario proposed by \citet{GLS02} consists of the formation
of a transient binary, which becomes bound with the aid of dynamical friction
from the surrounding sea of small bodies ($L^2_s$ mechanism,
\S\ref{sec-dfmech}). \citet{SR07} demonstrated that at the distance of the
Kuiper belt, $L^2_s$ dominates over $L^3$. \cite{A05} and \cite{laf07} suggest
that transient binaries that spend a long time in their mutual Hill sphere,
near a periodic orbit, form the binaries in the $L^2_s$ and $L^3$
mechanisms. \citet{SR07} investigated the relative importance of these
long-lived transient binaries. They found that such transient binaries are not
important for binary formation via the $L^3$ mechanism and that they become
important only for very weak dynamical friction in the $L^2s$
mechanism. \citet{F04} proposed a binary formation mechanism which involves a
collision between two large KBOs that creates a small moon. An exchange
reaction replaces the moon with a massive body with high eccentricity and
large semi-major axis.

Finally, \citet{nyr10} suggested that Kuiper belt binaries could form directly
during a gravitational collapse that leads to the formation of 100-km-sized
KBOs. This binary formation scenario is fundamentally different in the sense
that it assumes, based on recent work on the streaming instability
\citep{yg05,jom+07}, that 100-km sized KBOs formed by direct gravitational
collapse rather than coagulation. Whether such a gravitational collapse KBO
formation scenario is feasible and whether it can explain the observed KBO
size distribution, which is well matched by coagulation simulations
\citep{kl99,K02,kb04}, remains to be determined. Furthermore, a
gravitational collapse binary formation scenario, would have to provide a
convincing explanation for the large binary fraction of similar brightness
systems in the Kuiper belt and the absence of such binaries in the Asteroid
belt.  \cite{nyr10} appeal to enhanced binary disruption in the Asteroid belt
due to collisions and scattering events to explain this difference.  A
dynamical origin, which is common in all the other binary formation scenarios
discussed above, takes advantage of the fact that the Hill radius is more than
an order of magnitude larger for KBOs compared to similar sized objects in the
Asteroid belt. Gravitational capture binary formation scenarios therfore
support the observations that the Kuiper belt is the ideal place for wide,
similar brightness binaries to form.

We decided to focus on the \citet{GLS02} binary formation mechanism for the
calculations in this paper. Our choice is motivated by the success this binary
formation scenario has had in explaining the binary abundance and the observed
binary separation distribution. It also predicts the existence of triplet
systems. The first triplet system was reported recently by \citet{B10}. One
set of observations that are not successfully explained by the \citet{GLS02}
formation scenario is the mutual binary inclination distribution
\citep{N08}. Binary formation rates are extremely sensitive to the velocity
dispersion $v \sim \left<e\right>\Omega a$ of the population of large
planetesimals from which they form, where $\left<e\right>$ is the typical
eccentricity of a planetesimal.  Efficient binary formation requires that the
large KBOs have a velocity dispersion that is less than their Hill velocity,
$v_H = \Omega R_H$.  Binary formation rates quickly exceed the age of the
solar system, once the velocity dispersion grows significantly above $v_H$
\citep{N08,SR07}. Binary formation while sub-Hill velocities prevailed,
however, predicts low mutual binary inclinations \citep{clm+07,SR08} and is
inconsistent with the roughly uniform inclination distribution that is
observed (see \citet{N10} and references theirin).  Evolution of the mutual
binary inclination after formation provides a possible solution to this
problem. Although it has, for example, been suggested that the Kozai mechanism
could affect the orbital evolution of Kuiper belt binaries \citep{P09}, it
remains a topic of ongoing research if mutual binary inclinations can be
excited to sufficiently large values that would allow such a mechanism to
operate.

\subsection{Binary Capture Mediated by Dynamical Friction}\label{sec-dfmech}

After specifying our assumptions, we now review the binary formation rate for
the $L^2s$ mechanism.  %Along the way, we quote a few intermediate expressions
which will be useful in calculating the expected binary fraction.  Following
\citet{GLS02}, we use the ``two-group approximation", which consists of the
identification of two groups of objects, small ones, that contain most of the
total mass with surface mass density $\sigma$, and large ones, that contain
only a small fraction of the total mass with surface mass density $\Sigma \ll
\sigma $. We assume that $\sigma \sim \sigma_{40} (a/40~AU)^{\gamma}$ which is
a generalization of the minimum-mass solar nebula \citep{w77,H81}, where
$\sigma_{40} \sim 0.3 \;\rm{ g~cm^{-2}}$ is the surface density at a
heliocentric distance of $40\rm{AU}$.  The power-law index $\gamma$ is
typically assumed to have values between -1 and -1.5, with submillimeter
observations of the outer regions of protoplanetary disks favoring $\gamma
\approx -1$ \citep{awh+09}.

%We are ultimately interested in the scaling of binary abundance as a function
%of semi-major axis from the Sun (Section \ref{sec-abund}).  This scaling will
%depend on $\gamma$, but not on $\sigma_{40}$---for our purposes, the value of
%$\sigma_{40}$ only matters in that it sets the regime in which binary
%formation occurs.  

We likewise parameterize the mass
surface density of large bodies with sizes of $ R \sim 100~\rm{km}$ and larger
as $\Sigma \sim \Sigma_{40}(a/40~AU)^{\beta}$, where we
treat $\beta$ as a free parameter.  Estimates from Kuiper belt surveys
\citep[e.g.,][]{TJL01,TB03,PK08,FK09,fgh09} yield $\Sigma_{40} \sim 3 \times
10^{-4} \; \rm{g~cm^{-2}}$ in the current belt.  

%We will show that $\Sigma$ does not explicitely enter our estimate for the binary abundance---again, it only matters in setting the regime for binary formation.

%,  and we assume that $\Sigma$ during
%the formation of Kuiper belt binaries was the same as it is now. Our choice
%for $\Sigma$ and $\sigma$ is also consistent with results from numerical
%coagulation simulations \citep[e.g.,][]{kl99,K02,kb04}.

\citet{GLS02} demonstrate that, under the assumption that $\Sigma$ was
the same as it is now during the formation of Kuiper belt binaries, the
velocity dispersion $v$ of $\sim$$100\rm{km}$ KBOs is damped by dynamical
friction from the sea of small bodies such that $v < v_H$.  This is referred
to as the ``shear-dominated velocity regime" because under these conditions,
the relative velocity of two large bodies that encounter one another is
dominated by their Keplerian shear.  Binary formation is inefficient when $v
\gtrsim v_H$ \citep{clm+07,SR08}.

In particular, in this scenario, large bodies grow by the accretion of small bodies. Large KBOs viscously stir the small bodies.  This stirring increases the small bodies' velocity dispersion, $u$,
%, their velocity dispersion $u$ 
which grows on the same timescale as $R$ and is given by
\begin{equation}\label{e2}
\frac{u}{v_H} \sim \left( \frac{\Sigma}{\sigma \alpha} \right)^{1/2} \sim 3
\left ( \frac{a}{40~AU} \right)^ {(\beta-\gamma+1)/2}
\end{equation}
where $\alpha = R/R_{H}$ \citep{GLS02}.\footnote{Collisions among the small bodies and the associated collisional damping of
their velocity dispersion is unimportant on the viscous stirring timescale and
binary formation timescale, provided that the small bodies have radii of order 100m or larger.} Meanwhile, the velocity $v$ of large KBOs increases due to mutual
viscous stirring, but is damped by dynamical friction from the sea of small
bodies such that $v < u$. Balancing the stirring and damping rates of $v$ and
substituting for $u$ from Equation \ref{e2}, we find
\begin{equation}\label{e3}
\frac{v}{v_H} \sim \alpha^{-2} \left(\frac{\Sigma}{\sigma}\right)^{3} \sim
0.1 \left( \frac{a}{40~AU} \right)^{3(\beta-\gamma)+2}. 
\end{equation} 
%For our parameters, we therefore have that $v < v_{H}$ during the epoch of
%formation of bodies with sizes of $\sim 100\rm{km}$ and their binary
%formation.  The condition $v < v_H$ is referred to as the ``shear-dominated velocity regime" because under these conditions, the relative velocity of two large bodies that encounter one another is dominated by their Keplerian shear.  Binary formation is inefficient when $v \gtrsim v_H$.

A transient binary forms when two large KBOs penetrate each other's Hill
sphere. This transient binary must lose energy in order to become
gravitationally bound.  In the $L^2_s$ mechanism, the excess energy is carried
away by dynamical friction with the sea of small bodies.  Since it is not
feasible to examine the interactions with each small body individually, their
net effect is modeled by an averaged force which acts to damp the large KBOs'
non-circular velocity. We parameterize the strength of this damping by a
dimensionless quantity $D$ defined as the fractional decrease in non-circular
velocity due to dynamical friction over time $\Omega^{-1}$:
\begin{equation}\label{e5}
D\sim \frac{ \sigma}{\rho R} \left(\frac{u}{v_H}\right)^{-4} \alpha^{-2} \sim
\frac{\Sigma}{\rho R} \alpha^{-2} \left(\frac{v}{v_H}\right)^{-1} \;\;,
\end{equation} 
where $\rho \approx 1$ g/cm$^3$ is the internal density of a KBO.  The first expression is an estimate of dynamical friction by the sea of
small bodies for $u>v_H$. The second expression describes the mutual
excitation among the large KBOs for $v<v_H$. These two expressions can be
equated since the stirring among the large KBOs is balanced by the damping due
to dynamical friction. Using this parameterization for the dynamical friction,
the binary formation rate for equal mass bodies via the $L^2s$ mechanism in the
shear-dominated velocity regime is given by
\begin{equation}\label{e6} 
FR_{L^2s}=A_{L^2s} D \left(\frac{\Sigma}{\rho R}\right) \alpha^{-2} \Omega
\end{equation} 
\citep{GLS02} where $A_{L^2s}$ is a constant with a value of about 1.4
 \citep{SR07}. 
 %We have used the fact that in the shear-dominated
%velocity regime, the growth of inclinations is suppressed and the disk of KBOs is effectively two-dimensional
%\citep{WS1993,R2003,gls04}. 
 $FR_{L^2s} \sim 1.3 \times 10^{-5} \rm{yr^{-1}}$ when evaluated
 at 40~AU.

\subsection{Binary abundances}\label{sec-abund}

We now calculate the expected variation in the fraction of binaries formed as a function of semi-major axis.  The binary separation, $s$, shrinks due to dynamical friction at a rate
$s^{-1} |ds/dt|$. The probability of finding a large KBO in a binary per
logarithmic band in $s$ is given by
\begin{equation}\label{e8}
s~ p(s) \sim \frac{FR_{L^2s}}{s^{-1} |ds/dt|}
\end{equation}
where $p(s)$ is the probability density of finding a large KBO in a binary
with binary semi-major axis $s$. There are two regimes in binary separation
that need to be considered separately. In the first regime the binary
semi-major axis is sufficiently large such that the velocity of the binary
components around their common center of mass, $v_B$, is small compared to the
velocity dispersion of the small bodies that provide the dynamical friction,
i.e. $v_B < u$. In this case $s^{-1} |ds/dt| = D$, given by Equation
(\ref{e5}). However this regime only applies to binary semi-major axes that
satisfy $ GM/u^2 < s < R_H$ which translates into binary angular separations
of 3'' and greater at heliocentric distances of about 40~AU. To our knowledge,
only one \citep[2001 QW322,][]{P08} out of about 70 currently known KBO
binaries satisfies this criterion. We therefore focus from here onwards on the
second regime for which $v_B > u$, which applies to binaries with angular
separation of less than about 3'' at a distances of about 40~AU. In this regime, the binary separation
shrinks due to dynamical friction at a rate
\begin{equation}\label{e9}
\frac{1}{s} \left| \frac{ds}{dt} \right| \sim \frac{ \sigma}{\rho R}
\left(\frac{u}{v_H}\right)^{-2} \left(\frac{v_B}{v_H}\right)^{-2} \alpha^{-2}
\Omega.
\end{equation}
Substituting the above expression and the binary formation rate from Equation
(\ref{e6}) into Equation (\ref{e8}) we find
\begin{equation}\label{e10}
s~ p(s) \sim \frac{ \Sigma}{\rho R} \left(\frac{u}{v_B}\right)^{-2}
\alpha^{-2} \sim \frac{ \sigma}{\rho R} \left(\frac{R}{s}\right) \alpha^{-2}
\end{equation}
 where in the last step, we used the expression for $u$ given in Equation (\ref{e2}) \citep{GLS02}. The only semi-major axis dependence that enters in Equation (\ref{e10}) results from
 $\sigma$ and $\alpha$. For $\sigma \propto a^{\gamma}$ the binary fraction
 as a function of $a$ is given by
\begin{equation}\label{e11}
s~ p(s) \sim \frac{ \sigma}{\rho R} \left(\frac{R}{s}\right) \alpha^{-2}
\propto a^{\gamma+2}
\end{equation}
at the time of formation.  

Equations (\ref{e10}) and (\ref{e11}) show that the binary fraction does not
explicitely depend on $\Sigma$. We also note that, because we are interested in
the scaling of $s~ p(s)$ with $a$, the values of $\sigma_{40}$ and
$\Sigma_{40}$ do not matter as long as their ratio is such that binary
formation can proceed in the shear dominated velocity regime. Given everything
else equal, we find that the binary fraction is between 1.4 times
($\gamma=-3/2$) to twice ($\gamma=-1$) as high at 50~AU compared to 25~AU
during the epoch of binary formation. This trend in the binary fraction as a
function of heliocentric distance is in agreement with observations that find
a binary fraction that is more than a factor of 2 lower in the excited and hot
classical populations, which most likely formed closer to the Sun and were
scattered to their current locations, than in the cold classical population,
which most likely formed in-situ at about 40~AU.  An analogous calculation for
the $L^3$ mechanism of binary formation yields the scaling relation $s~ p(s)
\propto a^{\gamma-1/4}$, where we used $\Sigma/\sigma \sim \alpha^{3/4}$
\citep{SR11}.

In Section \ref{sec-binfrac}, we explore the expected binary fraction as a
function of current dynamical class in more detail.  We recall here that the
power-law index of the small bodies' mass surface density, $\gamma$, is
uncertain. Often $\gamma = -1.5$ is used based on the minimum mass solar
nebula \citep{H81}, but protoplanetary disk observations at distances
comparable to the Kuiper belt \citep{awh+09} suggest $\gamma = -1$.  In the
calculations that follow, we assume $\gamma = -1$. If $\gamma = -1.5$ then the
trends in the binary abundance will still be present but smaller in magnitude.
Particle pile-ups \citep[e.g.,][]{yc04}, if they occur, result in a shallower
mass surface density profile, making trends in binary abundance more
pronounced.

Finally, as mentioned above, efficient binary formation (as proposed by
\citet{W02,GLS02,F04,A05,laf07}) requires $v \lesssim v_H$, which implies that
the velocity dispersion of the large bodies that form binaries has to be
damped to sub Hill velocities. If this damping is provided by dynamical
friction generated by small bodies, as we have assumed in deriving Equation
(\ref{e3}), than this implies that $\Sigma \ll \sigma$, because $v>v_H$
otherwise. These binary formation scenarios are therefore inconsistent with
reduction of the mass in the Kuiper-belt by two orders of magnitude or more in
an entirely size-independent fashion, for example in one large scattering
event. Analytic work and numerical coagulation simulations that model the
growth of KBOs indeed find that only about $\alpha^{3/4} \sim 10^{-3}$ of the
initial disk mass is converted into large KBOs suggesting that $\Sigma/\sigma
\sim 10^{-3}$ \citep{kl99,kb04,SR11}, which is consistent with the surface
density values used in this section and the conclusion that $v<v_H$.

\section{KBO BINARY FRACTIONS AS A FUNCTION OF DYNAMICAL CLASS}\label{sec-binfrac}

In order to test the nature of Neptune's migration with the abundance of
comparable-mass, wide-separation KBO binaries, we suggest the following procedure.  We recall that we are searching for a signature of an outer solar system history in which the cold (low inclination) classical population formed in situ, the hot (high inclination population) formed closer to the Sun and was dynamically mixed during transport to its current location, and resonant KBOs are a superposition of these two populations, caught into resonance during migration of Neptune spanning between several and $\sim$10 AU (\S\ref{sec-stage}).  Our proposed signature of this class of migration histories is summarized in Figure \ref{fig-fracts}.

\subsection{Looking for a Cold Component}\label{sec-cold}
First one needs to
establish whether a low inclination component, analogous to the cold classical
population, exists in a given resonant population.  Recall that because any hot population will contain low-inclination bodies, a cold component refers not merely to objects with low-inclinations but rather to a separate low-inclination population.  As yet, for the 3:2 there is no
dynamical evidence for two inclination components, implying that if one exists it is less pronounced than in the classical belt, and insufficient data is available for the 2:1 resonance to make this determination \citep{gea+10}. 
We note that because 3:2 resonant KBOs originated in a different location in the disk than classical KBOs if migration-induced capture occurred, and because the inclinations of resonant bodies evolve more over long-term integrations than their classical counterparts (Volk \& Malhotra, in prep.), the characteristic inclination that 
divides the low and high inclination populations in the 3:2 resonance could be different from that in the
classical belt.  Furthermore, this characteristic inclination could vary between resonances.
If a resonance does contain a cold
component with a broader width or a smaller relative contribution than seen in the classical belt, it should be
observable in binaries, just as the cold component can be picked out just from the
binary fraction in the classical belt \citep{ngs+08}.
We therefore suggest searching for a cold component by comparing the binary fractions of low and high inclination objects in a given resonance.  We discuss this distinction in the context of KBO colors and sizes in Section \ref{sec-colors}.
 
Theoretically, if cold and hot components are present in a resonance, our schematic migration scenario predicts fewer binaries in the high inclination population.  This is both because they originated
closer to the Sun and because the excitation process may have broken binaries
that were originally there.  \citet{pk10} have shown numerically that binary disruption is common for objects transported across the semi-major axis of Neptune.  They consider the particular chaotic capture scenario modeled in \citet{lmv+08} in which all KBOs, including those ultimately on low eccentricity and low inclination orbits, begin interior to the current cold classical belt, and they find that wide binaries are efficiently destroyed.  This finding is in conflict with the observed large binary fraction in the cold classical belt \citep{N08}.  \citet{pk10} argue that the difference in binary fraction between the cold and hot classicals may reflect scattering of the hot classicals off of Neptune during their history, contrasted with a gentler dynamical history for low inclination objects, which formed exterior to Neptune and never encountered the planet.
Depending on the details of the as yet unknown process exciting inclinations in the Kuiper belt, there
may be trends with inclination in the binary fraction of the high inclination population.  Nevertheless, we expect the
overall fraction to be smaller than that in the low inclination population.

A lack of a cold component in the 3:2 resonance might be explained in the context of migration-induced capture if the original disk was substantially excited interior to the current location of the 3:2, though no proposed model currently produces such excitation.  In contrast, a lack of a cold component in the 2:1 resonance would
seriously challenge a scenario invoking smooth migration through a hot plus cold
planetesimal disk because the 2:1 resonance should have swept through the
location of the current cold classical belt.

We emphasize that the 3:2 and 2:1 resonances are the best locations to test for the presence of
a cold component. Higher order
resonances preferentially capture objects with large
eccentricities. Therefore, if eccentricities and inclinations are correlated,
higher order resonances may only contain hot components even if they swept
through a disk that had a cold component. However, if a higher order resonance
does have a population of low inclination objects, it should also have a higher binary fraction
in the cold component compared to the hot component.

\subsection{Looking for Trends in the Low Inclination Resonant Population}
Provided that a cold component can be established in a Kuiper belt resonance
(Section \ref{sec-cold}), we suggest looking for the following trends in binary fraction within the low inclination resonant population.  First, we propose looking for differences across the binary fractions of the cold populations of different resonances and of the cold classical belt, and second we propose looking for trends in binary fraction with eccentricity within the low inclination population of a single resonance.  Because it requires observations for fewer objects per resonance, the former effect may be easier to measure, and we describe it first.

\begin{figure}
%\plotone{fractions_bw.pdf}
%\plotone{f3.eps}
%\plotone{f3.pdf}
%\epsscale{0.5}
\centering
\includegraphics[scale=0.6]{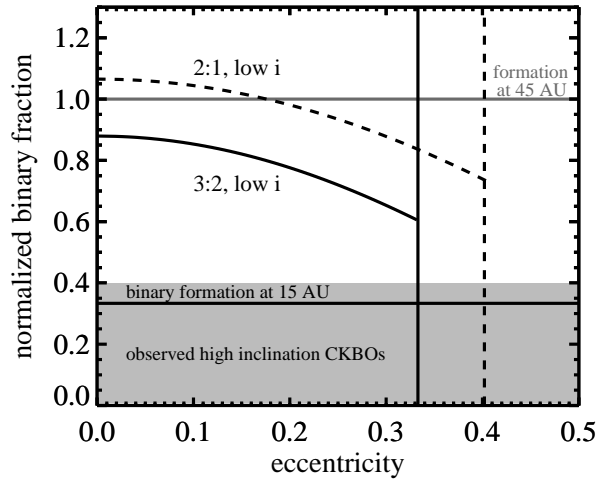}
\caption{Fraction of low-inclination, resonant Kuiper belt objects predicted to be wide-separation, roughly equal-mass binaries, under the hypothesis that Neptune captured objects into resonance as it migrated $\sim$10 AU.  We normalize this expected fraction to the binary fraction among low-inclination (cold)  classical objects (CKBOs) at 45 AU.  The vertical lines show both the maximum eccentricity excitation from migration by 9.4 AU (calculated using Equation \ref{eqn-aeinvar}) and the maximum observed eccentricities in 3:2 and 2:1 resonances.  The agreement of these values is promising for the migration scenario, but chaos at large eccentricities might also explain the different maximum eccentricities in different resonances.  The observed binary fraction among highly inclined (hot) CKBOs (shaded region) is consistent with formation $\lesssim$15 AU from the Sun, though scattering events could disrupt binaries, complicating this interpretation.
}
\label{fig-fracts}
\end{figure}  

Figure \ref{fig-fracts} displays the binary fraction as a function of eccentricity for low inclination objects in the 3:2 and 2:1 resonances, normalized to the binary fraction of cold classical KBOs at 45 AU.  These relative fractions are calculated by combining the expected formation locations of the resonant KBOs from Equation (\ref{eqn-ai}) with the trend in binary formation rate as a function of formation location given by Equation (\ref{e11}) and $\gamma = -1$.  The difference between true and expected formation location is discussed in Section \ref{sec-varymig}.

In general, the larger the semi-major axis of a resonance, the higher the binary fraction of its cold
component should be. To zeroth order, we
expect the binary fraction in the 2:1 resonance to be roughly comparable to the cold
classicals.  A higher fraction of large eccentricity objects in resonance
would lower this estimate, as can be seen from Figure \ref{fig-fracts}. To
correct for this bias in a given observed sample, one could, for example,
calculate the estimated fraction using the eccentricity of each observed
object from Figure \ref{fig-fracts} and calculate the average. In contrast, we expect cold
3:2 objects to have a 20--30\% lower binary fraction than the cold classicals.  
Again,
for a particular set of observations, one can calculate an expected fraction as
above.  

Uncertainties in the population statistics of resonant KBOs prevent us from making a robust estimate of the number of measurements required to distinguish this effect.  Nevertheless, we provide the likely scale of the observations required by calculating the following example.  If the binary fraction of cold classicals at 45 AU is $\sim$30\%, consistent with the observed binary fraction among cold CKBOs \citep{ngs+08}, then averaged over all eccentricities, the low inclination 3:2 objects are predicted to have binary fraction $\sim$0.22, compared with $\sim$0.27 in the 2:1 resonance and $\sim$0.29 in the cold classical belt.
Identifying such subtle differences would require measuring the fraction of each population to within approximately 2--3\%.  \citet{ngs+08} measure a binary fraction of $29^{+7}_{-6}$\% in the cold classicals based on binary detection for 17 of 58 targets.  We estimate that a sample containing approximately $N\sim 250$ objects would generate errors of $\sqrt{0.3 N}/N = 0.035$, enough to distinguish our predicted binary fractions for cold 3:2 objects from the cold classicals.  This approximate number of observations would be required in each population.  The binary fractions in the cold 3:2 and cold 2:1 resonant populations could be separated with samples of $\sim$500 objects each. 

Currently, only 101 3:2 objects are known (as defined in \S\ref{sec-varymig}), 54 of which have $i < 10^\circ$ and 26 of which have $i < 5^\circ$.  In the 2:1 resonance, 15 objects are known, 9 of which have $i < 10^\circ$ and 6 of which have $i < 5^\circ$.  However, according to \citet{kjg+09}, approximately $2.5^{+2.5}_{-1.2}\times 10^4$ Plutinos are present in the belt with solar system absolute magnitude in the $g^\prime$ filter of $H_g < 10$.  Given
\begin{equation}\label{eqn-diam}
D(\rm km) = 2\times 1.5\times 10^8 p^{-0.5} 10^{0.2(m_\odot-H)} 
\end{equation}
\citep[e.g.,][]{pkg+08} with $m_\odot = -26.47$ in $g^\prime$ \citep{kjg+09},  $H = H_g < 10$ corresponds to diameters $D > 50$ km.  We have used albedo $p = 0.1$, consistent with observed KBOs in the visible \citep{bgs+09}.  

We are interested in objects near the break in the KBO size distribution, corresponding to the largest number of KBOs that are not likely to have collisionally evolved over the age of the solar system.  This break is measured at apparent magnitude $\sim 25$ in $R$-band \citep{fgh09}.   At a distance $d$ from the Sun, an object with absolute magnitude $H$ has apparent magnitude $H + 2.5\log(d^2\Delta^2/{\rm AU}^4)$, with $\Delta \approx d - 1$ AU.  Using $d = 42$ AU and $m_\odot = -27.12$ (appropriate for the Bessel R filter\footnote{{\tt http://mips.as.arizona.edu/\~{}cnaw/sun.html}, August 2010.}), 
%\label{magnote}
the break is at $D \sim 60$km.  This differs from the 90 km diameter quoted in \citet{fgh09} due to our larger assumed albedo.  The actual diameter at the break is uncertain to within approximately a factor of 2.  Since $D\sim50$km equals the size at the break to within the errors, we conservatively adopt a number of Plutinos equal to half of the estimated number of Plutinos with $H_g < 10$.
Given these assumptions, if a cold population of 3:2 KBOs is shown to exist with $i < 10^\circ$  and if the fraction of Plutinos with $i < 10^\circ$ is $\sim$0.5 as in the currently observed sample, then the 3:2 population likely contains at least $\sim$$6\times10^3$ dynamically cold objects larger than the break in the size distribution. More than an order of magnitude more Plutinos are likely present in the belt than would be required to compare the binary fractions of cold 3:2 objects and cold classicals.  

The number of objects in the 2:1 resonance is less well constrained than the number in the 3:2 resonance.  \citet{cj02} estimate that the 2:1 hosts a factor of $\sim$3 fewer objects than the 3:2, and Lawler et al.~(in prep.) estimate a factor of $\sim$4 fewer objects in the 2:1 than in the 3:2 based on the well-characterized Canada-France Ecliptic Plane Survey.  Given a factor of 4 reduction and assuming, in analogy with the 3:2, that half of the 2:1 objects are members of a cold population, a factor of $\sim$3 more 2:1 objects are present than we predict are required to differentiate between the binary fractions of the cold 3:2 and 2:1 populations.

LSST is currently projected to detect more than 30,000 KBOs with diameters larger than $\sim$100 km \citep{ita+08}.  Using $m_\odot = -27.08$ (SDSS $r^\prime$ filter\footnotemark[8])
%\ref{magnote}
and albedo $p = 0.1$, the LSST limiting magnitude of $\sim 24.5$ in $r$-band corresponds to
 KBOs with diameters  $D \gtrsim$ 70 km at the semi-major axis of the 3:2 resonance ($d \approx 39$ AU) and $D\gtrsim100$ km at the semi-major axis of the 2:1 resonance ($d \approx 48$ AU).  In practice, smaller resonant KBOs will be observable because the large eccentricities of these objects bring them closer to the Sun.
With an angular resolution of $\sim$0.7" \citep{ita+08}, LSST will likely be able to distinguish some wide-separation binaries without requiring follow-up observations.  However, the observations in \citet{ngs+08} using the Hubble Space Telescope have a factor of 10 better resolution and include binaries that will not be resolved by LSST.  \citet{kjg+09} estimate that Plutinos are $\sim$20\% as abundant as CKBOs.  If only $\sim$2\% or more of the objects discovered by LSST are Plutinos and if the fraction of wide binaries among these objects is measured, then LSST will detect sufficiently many low inclination Plutinos to search for our predicted difference between the binary fractions of cold 3:2 and cold classical objects. 

LSST may also detect the $\sim$500 low inclination 2:1 objects required to differentiate between the 3:2 and 2:1 populations.  Conservatively choosing a size distribution $dN/dR \propto R^{-q}$ with $q = 5$ \citep[at the large end of the $q=2$--5 range measured for various subpopulations of the belt,][]{fgh09,FK09,fbs10}, we estimate the number of cold 2:1 objects with $D\sim 100$km to be $\sim$200.  These objects will be observable by LSST at their semi-major axis.  A KBO in 2:1 resonance with $e = 0.2$ spends $\sim$16\% of its time at distances from the Sun of no more than 40 AU.  At these distances, objects with $D\sim 70$ km will be visible, adding another $\sim 800\times0.16\sim100$ observable low-inclination bodies, for a total of $\sim$300 objects.  Given the substantial uncertainties encompassed in this calculation and our conservative choice of $q = 5$, it is plausible that LSST will detect sufficient low inclination 2:1 bodies to differentiate between the cold 3:2 and 2:1 populations.

We now turn to the search for trends in binary fraction with eccentricity within the low inclination population of a single resonance.
As illustrated in Figure \ref{fig-fracts}, within a single resonance, the binary fraction at small eccentricities is $\sim$40\% larger than that at the highest observed eccentricities.  If we consider the average of all objects with $e < 0.2$ compared with those having $e \ge 0.2$, the difference is reduced to $\sim$20\%, though this value depends on the distribution of eccentricities included in the average.  

Extending our example calculation above, we estimate that low inclination 3:2 objects have binary fraction $\sim$0.25 for $e < 0.2$ and $\sim$0.21 for $e \ge 0.2$.  
These fractions could be separated with samples of 
$N\sim 750$ in each of the $e < 0.2$ and $e \ge 0.2$ subsets of the low-inclination data.  
Again, these numbers depend on the distribution of eccentricities of the objects in the resonances.  Though the observational effort needed would be substantial, this experiment should be possible using KBOs discovered by LSST.
An order of magnitude more objects are likely present in the belt than would be required to test this effect, and LSST will detect a sufficient number of Plutinos to search for our predicted trend if $\sim$10\% or more of its projected KBO discoveries are in the 3:2 resonance.

In our example, the low inclination 2:1 objects are predicted to have binary fraction $\sim$0.31 for $e < 0.2$ and $\sim$0.26 for $e \ge 0.2$.  Separation of these fractions requires $\sim$1000 low-inclination 2:1 objects, equally divided between low and high eccentricities.  We estimate that enough objects exist in the belt to perform this experiment.  Though our estimate indicates that LSST will discover a factor of $\sim$3 fewer 2:1 KBOs than required, we emphasize that this calculation employs a conservative size distribution and is highly uncertain because the intrinsic population of 100 km objects in the 2:1 resonance is not well known.

We recall that because, in the migration scenario, the 2:1 resonance passed through the cold classical belt, the 2:1 resonance should provide the clearest case for our dynamical signature.   If the 3:2 resonance has a cold component, then it should also provide a good case for
testing the binary abundance. As mentioned above, higher order resonances are
less likely to capture unexcited objects. However, if they have some cold
component, then they should show a similar trend in the binary abundance as a
function of eccentricity as the 3:2 and 2:1 resonances.

We re-emphasize that these trends should only exist among an observed
cold (low-inclination) component of the resonant population, if such a component exists.  We further emphasize that we propose to test the binary fraction and not the absolute number of binaries in the resonant populations. Long-term differences in the loss rates from different resonances or as a function of orbital parameters within any given resonance \citep[e.g.,][]{tm09} will not affect our results because these losses should affect binaries and single KBOs equally.
Finally, we note that
the KBOs need to be tightly in resonance (small libration angle) rather than
tenuously in resonance, since they may otherwise have been transiently captured
by scattering \citep{lm06}.

\subsection{Interpretation for Different Dynamical Sculpting Models}

If a cold component is observed in the fraction of wide-separation, comparable mass binaries in one or more resonances, it will provide support for migration-induced capture, which naturally produces a cold component within resonances.  Migration-induced capture may be consistent with no cold component in the 3:2 resonance, 
but the lack of a cold component in the 2:1 resonance would argue strongly against this model.  In contrast, in chaotic capture as envisioned by \citet{lmv+08}, 
the binary fraction in the resonances should match the fraction in the scattered disk, even at low inclinations.  Given the presence of a distinct binary fraction in the cold classical disk, for chaotic capture to be viable, the models presented in \citet{lmv+08} need to be adjusted to allow maintenance of a cold classical population that formed {\it in situ}.  Even if such a modification is possible, it would not generate a cold component within the resonances, retaining the search for a cold resonant component as a useful distinguishing test.

Measurement of a trend in binary fraction with eccentricity within the cold component of a resonance would provide conclusive evidence of migration-induced capture.
Unfortunately, lack of such a trend would not constitute conclusive evidence that migration-induced capture did not occur. We expect such a trend to be a subtle effect, and uncertainty in the surface-density profile of the initial disk means that we cannot conclusively predict its magnitude.  Lack of an observed trend could imply a steeper surface density (nearer $\gamma = -2$) or a different binary formation scenario.
Similarly, if a trend in binary fraction with eccentricity is observed among low-inclination resonant KBOs, but the trend differs from our prediction, this could imply either a different surface density profile or an alternative 
binary formation mechanism. 

\section{KBO COLORS AND SIZES}\label{sec-colors}

Binary fraction is not the only physical characteristic that has been shown to differ between dynamical populations in the Kuiper belt---striking correlations with KBO inclination have also been measured for KBO colors \citep{tr00,tb02,PL08} and sizes \citep{ls01a}.  These characteristics are more difficult to interpret than the binary fraction because no quantitative theory exists for how they should vary as a function of formation location.  Nevertheless, we now consider the procedure outlined in Section \ref{sec-binfrac} in light of these properties.  We address the 3:2 resonance, where data is most abundant.  A detailed reassessment of the available data is warranted but beyond the scope of this paper.  For our purposes, we demonstrate that currently measured size and color distributions, as compiled by \citep{apc09} for 3:2 objects and by  \citep{PL08} for classical objects, do not conclusively distinguish between our two dynamical scenarios.

\begin{figure}[ht]
%\epsscale{0.4}
%\plotone{iBRD.pdf}
%\plotone{f4.eps}
%\plotone{f4.pdf}
\vspace{-0.2in}
\includegraphics[scale=0.75]{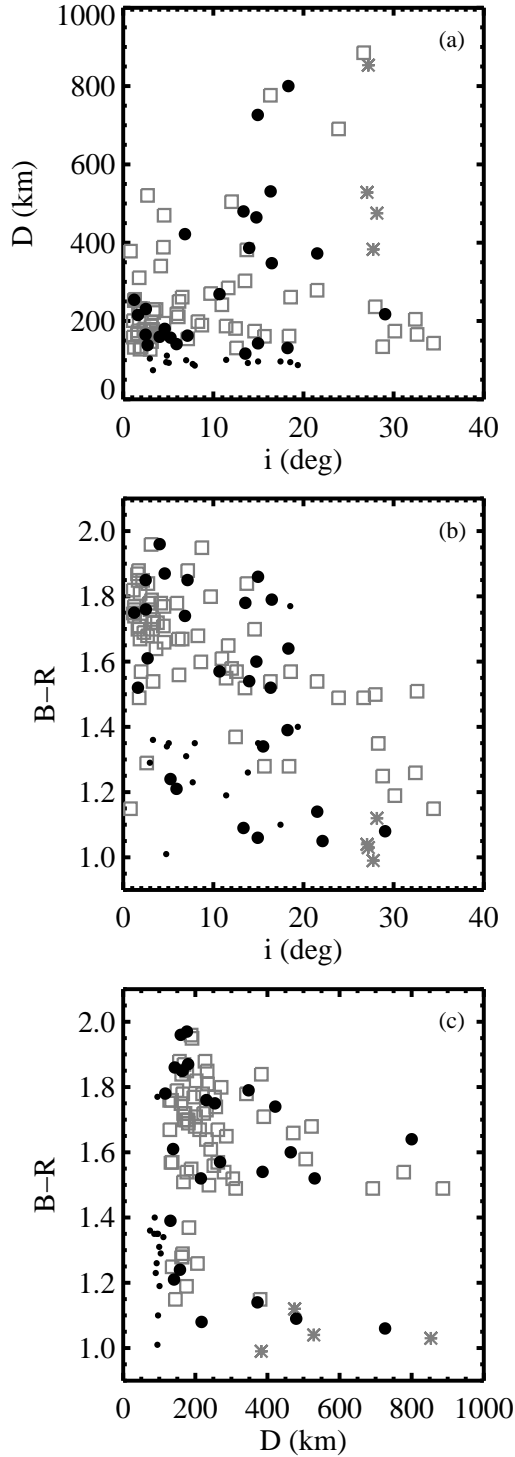}
\caption{Observed diameter as a function of inclination with respect to the invariable plane (a), inclination vs.~B-R color (b), and diameter vs.~B-R color (c) for classical (gray squares) and 3:2 resonant (black circles) KBOs.  Small circles correspond to 3:2 objects with sizes smaller than the minimum-size classical KBO.  Classical objects that are known members of the Haumea collisional family (gray asterisks) have unusually blue colors, known to result from water ice features.  Data for resonant objects is compiled by \citet{apc09} and that for classical KBOs is compiled by \citet{PL08}.  See the text for details of our conversion from magnitudes, $H_R$, to diameters.}
\label{fig-colorsizeinc}
\end{figure}

First, does the 3:2 resonance contain a cold component?  Figures \ref{fig-colorsizeinc} (a) and (b) display KBO diameters and B-R colors as a function of inclination for objects in the 3:2 resonance and for classical objects.   We convert inclinations to be with respect to the invariable plane.  We calculate diameters from the absolute solar system magnitudes $H_R$ quoted in \citet{apc09} and \citet{PL08} using Equation (\ref{eqn-diam}) with $H = H_R$ and $m_\odot = -27.12$ (Bessel $R$ filter\footnotemark[8]).
%\ref{magnote}
In keeping with the high albedos observed for KBOs in the visible \citep{bgs+09}, we use albedo $p = 0.1$.  Members of the Haumea collisional family \citep{BB07,RB07,scd+10} are marked.  

Figure \ref{fig-colorsizeinc} (a) suggests that 3:2 objects with inclinations $< 5$--10$^\circ$ may have systematically smaller sizes than high inclination population, consistent with the trend seen among classical objects.  At first glance, the color trend is less promising.  However, as noted by \citet{apc09}, the smallest 3:2 resonant KBOs have systematically bluer colors than their larger compatriots (see Figure \ref{fig-colorsizeinc} (c)).  The smallest classical object in the sample considered here has $D\sim 110$ km, close to the observed break in the classical size distribution \citep{fgh09}.  We suggest that this color difference, unlike the color variation among large KBOs, may result from a collisional break in the population of 3:2 objects that is not yet observed among classical KBOs because colors have not been measured for small enough objects.  This possibility merits further investigation (Schlichting et al., in prep.).  \citet{apc09} alternatively suggest that this trend may come from collisions between Plutinos and Neptune Trojans.  If either idea is validated, only the colors of those KBOs too large to suffer ongoing collisions reflect conditions in the initial nebula and only those should be considered when searching for a difference in color between in a low-inclination and a high-inclination population.  Figure \ref{fig-colorsizeinc} (b) is suggestive, though certainly not conclusive, that such a difference exists, again divided by a characteristic inclination between 5 and 10$^\circ$.  This
is similar to the color break that \cite{PL08} find at $\sim$$12^\circ$
in the classical belt.  We highlight that the transition in physical
properties could be at $\sim$$10^\circ$ or even higher rather than at
$5^\circ$ as usually quoted, even in the classical belt.

Next, do 3:2 objects show a trend in size or color with eccentricity?  No such trend has been observed in the 3:2 population \citep{dbt+08}.  However, because no quantitative model exists for how large KBO colors vary with formation distance, it is not clear whether we would expect to see such a trend in colors.  For example, the difference in color between cold and hot classicals might result not from a continuous variation in the protoplanetary nebula but rather from critical transitions in the ability of KBOs to incorporate or retain different ices \citep[e.g.,][]{sb07}.  In this case, the critical transition separating the region of formation for hot and cold CKBOs could have been interior to the region swept by the 3:2 resonance, leaving no trend with eccentricity in the colors of cold resonant objects.  If such a trend does exist, we would expect higher eccentricity objects to have bluer colors.  
The number of large, low-inclination, 3:2 resonant objects is not large enough to determine whether 
this is the case.

\section{SUMMARY}\label{sec-sum}

Observations demonstrate that unlike asteroids, a substantial fraction of Kuiper belt objects are comparable-mass, wide separation binaries  \citep{N08}.  These bodies typically have sizes of $\sim$100 km.  Unlike the Pluto-Charon system and other large mass-ratio binaries, they likely formed by dynamical capture rather than through collisions.  Because the Hill radius of a planetesimal increases with distance from the Sun, the efficiency of dynamical capture is higher for binaries that formed farther from the Sun.

The binary fraction in the cold classical Kuiper belt is $\sim$29\%, compared with $\sim$10\% for hot classical KBOs \citep{ngs+08}.  This discrepancy suggests that objects in the hot classical belt formed closer to the Sun where binary formation was less efficient.  The binary fraction of the hot belt may also have been reduced by binary disruption during scattering interactions with Neptune \citep{pk10}.
In this paper, we have developed a method that uses the fraction of wide-separation, comparable mass binaries in various populations of the Kuiper belt to distinguish between two competing frameworks for the dynamical sculpting of the belt.  If current projections are correct, we expect LSST to discover sufficiently many Kuiper belt objects to perform our proposed tests.

In both dynamical scenarios, during the late stages of planet formation, Neptune was located $\sim$10~AU  or more closer to the Sun than its current semi-major axis of 30 AU.  A dynamical cataclysm excited the the planet's orbit, which was subsequently damped by dynamical friction with planetesimal debris.   
While Neptune's orbit was excited, it interacted with and excited the orbits of nearby planetesimal debris.  In this context, we consider the following 
two scenarios, named for the methods by which resonant KBOs are captured.
\begin{enumerate}
\item ``Migration-induced capture."  Neptune's orbit was damped onto an approximately circular, coplanar orbit while it still orbited $\sim$10 AU closer to the Sun than it does today.  It then migrated smoothly outward to its current location as it scattered planetesimal debris \citep{fi84}, capturing KBOs into resonance as it proceeded \citep{m93,m95}.  Each resonance captured objects from a distinct region of the protoplanetary disk, and the final eccentricity of a resonant KBO is correlated with the location of its capture into resonance.
\item ``Chaotic capture."  Neptune scattered quickly onto an orbit at most a few AU away from its current location.  The planet's large eccentricity produced a chaotic region of overlapping resonances extending out to the 2:1 resonance, allowing objects scattered by Neptune to fill the region of space currently containing the Kuiper belt.  Neptune's eccentricity then damped, leaving objects in resonance \citep{lmv+08}.  Due to the chaotic nature of this process, most or all information about the formation locations of KBOs was randomized.
\end{enumerate}

To distinguish between these models, we calculate the fraction of
wide-separation, roughly equal-mass binaries among KBOs with diameter
$\sim$100 km expected for different dynamical populations under the hypothesis
of migration-induced capture.  We assume that KBO binaries formed via
transient binary capture, made permanent by dynamical friction with small
bodies \citep[the $L^2s$ mechanism,][]{GLS02}.  We evaluate our expressions
for binary formation in a disk with surface density profile $\sigma \propto
a^{-1}$, consistent with observations of the outer regions of extrasolar
protoplanetary disks \citep{awh+09}.  Other binary formation mechanisms that
vary with location in the protoplanetary disk and other disk surface density
profiles will produce similar, but quantitatively different, signatures. In
particular, the $L^3$ mechanism produces an opposite trend in the binary
fraction with formation location compared to the $L^2s$ mechanism.
  
Figure \ref{fig-fracts} summarizes the pattern of binary fractions predicted as a result of migration-induced capture, given these choices.  A low-inclination component of the 2:1 resonance should exist with a larger binary fraction than its high inclination component.  Such a component may also exist in the 3:2 resonance.
Among bodies with inclinations less than 5--10$^\circ$, the fraction of binaries in the 2:1 resonance should be comparable to that in the classical belt, while the low-inclination component in the 3:2, if it exists, should have a fraction $\sim$20--30\% lower than among classical KBOs.  Within the low inclination component of the 3:2 or 2:1 resonance, objects with $e < 0.2$ should have a binary fraction $\sim$20\% higher than those with $e \gtrsim 0.2$.  Our calculations are not affected by dynamical loss processes that operate with varying efficiency in different resonant populations \citep[e.g.,][]{tm09}---these processes affect single and binary KBOs equally and we are interested only in the binary fraction. Chaotic capture does not produce these signatures.

Though measurement of the full range of dynamical signatures predicted in this paper will require a substantial observational effort, searches for low-inclination components in the binary fractions of the 3:2 and 2:1 resonances are within immediate reach.  When compared with the cold component observed in the classical belt, these cold resonant populations need not include the same fraction of the overall population, nor must they have the same characteristic inclination.  Lack of a low inclination component in the 2:1 resonance would indicate that migration-induced capture is unlikely.  Conversely, presence of a low inclination component in either resonance would constitute strong evidence for the migration-induced capture mechanism.  

\acknowledgements We thank Susan Benecchi and Keith Noll for helpful
discussions.  We thank Eugene Chiang, Renu Malhotra, and an anonymous referee
for comments on the manuscript.  R.M.-C. acknowledges support by an Institute
for Theory and Computation Fellowship at Harvard University.  H.S. is
supported by NASA through Hubble Fellowship Grant \# HST-HF-51281.01-A awarded
by the Space Telescope Science Institute, which is operated by the Association
of Universities for Research in Astronomy, Inc., for NASA, under contact NAS
5-26555. The simulation in Section \ref{sec-varymig} was run on the Odyssey
cluster supported by the FAS Sciences Division Research Computing Group.

%\bibliographystyle{apj}
%\bibliography{ruthsrefs_hilke}

\end{document}